# Doubly Metastable Synthesis for Accelerated Quantum Materials Discovery

Kyuho Lee[1*], Berit H. Goodge[2,3], Soonhyun Kwon[4], Yash Anand[1], Theo Manev[2], Maximilien F. Debbas[4], Weiliang Luo[5,6], Takehito Suzuki[7], Heather J. Kulik[5,6,8], Joseph G. Checkelsky[1*]

[1] *Department of Physics, Massachusetts Institute of Technology, Cambridge, MA 02139, USA*
[2] *Max Planck Institute for Chemical Physics of Solids, Nöthnitzer Str. 40, 01187 Dresden, Germany*
[3] *School of Applied and Engineering Physics, Cornell University, Ithaca, NY 14850, USA*
[4] *Department of Nuclear Science and Engineering, Massachusetts Institute of Technology, Cambridge, MA 02139, USA*
[5] *Department of Chemistry, Massachusetts Institute of Technology, Cambridge, MA 02139, USA*
[6] *Department of Chemical Engineering, Massachusetts Institute of Technology, Cambridge, MA 02139, USA*
[7] *Department of Physics, Toho University, Funabashi, Japan*
[8] *NSF Center for the Chemistry of Molecularly Optimized Networks, Duke University, Durham, NC 27710, USA*

## Abstract

New physics emerges from new materials, which in turn rely on the development of new synthesis techniques. Identifying the enlargement of metastable synthesis window as enabling for realizing novel materials, we develop diffusion-activated epitaxy, a technique which combines two metastable synthesis methods of diffusion and epitaxy. We observe that within a narrow window of investigated synthesis space we can experimentally identify the stabilization of four distinct phases, three of which show X-ray diffraction and scanning transmission electron microscopy characterization data that are incompatible with the crystal structures registered in experimental[1–4] and computational[4–10] inorganic structure databases. These results suggest an enlarged metastable synthesis window is enabled by this synthesis approach, identifying doubly metastable synthesis as a promising method for exploration of quantum materials.

[*]kyuho@mit.edu; checkelsky@mit.edu

Breakthroughs in condensed matter physics are often realized by the discovery of novel materials and their manipulations, enabling platforms for new, exotic physical phenomena.[11–15] The ability to identify and stabilize novel materials enables progress of this field, as well as more generally to the fields of materials science and chemistry. In addition to theoretical, computational, and characterization methods, a key component that drives materials discovery is the development and refinement of synthesis methods, as demonstrated for example by novel heterostructural semiconductors enabled by molecular beam epitaxy,[16] high-pressure phases including high-temperature hydride superconductors realized by diamond anvil cells achieving Mbar level of pressures,[17,18] two-dimensional single-atomic-layer materials and their twisted heterostructures synthesized by van der Waals exfoliation and twistronics methods,[19,20] and numerous metastable phases stabilized by electrochemical and soft-chemistry synthesis.[21,22] It has been observed that historically there is a doubling time of ~10 years[23] for reported inorganic structures; anecdotally this overlaps with the period of prolific development of synthesis methods that established many techniques currently used in laboratories (Fig. 1a).

One of the challenges towards this development of inorganic materials discovery is the growing complexity of the materials phase space as the focus shifts towards ternary and more complex materials systems (Fig. 1b). In addition to accelerated synthesis via automation,[24,25] another enabling direction is to expand the experimental capability to stabilize metastable materials. Proximate to the thermodynamically stable, global-minima phases in the free energy landscape lie numerous metastable, local-minima phases with higher formation energy.[26] These phases often promise wide variations in physical properties induced by their structural modifications[27] and, in some cases, can be stabilized using metastable techniques such as quench-cooling, intercalation, or epitaxy, which either kinetically trap the system at local minima or permanently adjust the thermodynamic parameters to manipulate the free energy landscape.[26,28,29]

Here, we propose a new regime of "doubly metastable synthesis" established via integration of existing metastable techniques. To demonstrate this new synthesis regime, we design a novel *diffusion-activated epitaxy* (DAE) technique which combines diffusion with epitaxy. This method exposes a higher-melting-point substrate into a lower-melting-point liquid bath of elements, akin to liquid epitaxy or bulk flux growth (Fig. 1c). The system is then exposed to an intermediate temperature range to activate the surface of the substrate to engage in diffusion reactions with the environment but at low enough temperature that the bulk of the substrate remains inert. This establishes a metastable environment at the substrate-liquid interface where diffusion occurs under an epitaxial framework.

## Diffusion-activated epitaxy (DAE)

Here, we demonstrate DAE with oxide substrates $LaAlO_3$ (111), $(LaAlO_3)_{0.3}(Sr_2TaAlO_6)_{0.7}$ (LSAT) (111), and $SrTiO_3$ (111). The synthesis process is similar to conventional bulk flux growth, except that these high-melting-temperature substrates are introduced to the crucible containing the flux and element powders. In short, the substrates are placed inside an alumina Canfield crucible with Sn flux and Co powders, all of which are vacuum-sealed inside a quartz tube and annealed with a typical temperature cycle operated for flux growth. This is followed by a quench-cooled centrifuging process to physically separate the liquid flux from the samples (see Methods). During high-temperature annealing, aluminum leaches out from the crucible, forming an Al-Co-Sn liquid environment at elevated temperatures. We first investigate whether the synthesis process involves epitaxy and substrate-environment interfacial diffusion as designed. Fig. 2a shows the X-ray diffraction (XRD) $\theta$-$2\theta$ symmetric scans of the three samples synthesized with a stoichiometric powder ratio of $CoSn_4$ (see Methods for detailed synthesis conditions). Each sample shows a distinct family of prominent Bragg peaks which manifest as bright localized spots in the XRD reciprocal space map (RSM) measurements (Fig. 2b-d), indicating the single-crystalline nature of the stabilized phases, with slight peak broadening suggesting a small distribution of misorientation. For the sample on LSAT, the in-plane reciprocal space component $q_x$ of the substrate and film peaks matches, indicating that the stabilized phase is strained to the substrate (Fig. 2c). Meanwhile, substrate- and film-peak $q_x$ are slightly mismatched for the samples on $LaAlO_3$ and $SrTiO_3$ (Fig. 2b,d), indicating that the stabilized phases on these substrates are slightly relaxed with respect to the substrate but still establish close in-plane lattice constant relationships with the respective substrates.

The phase on $SrTiO_3$ ($p_1$) is identified as an orthorhombic structure with $a \approx 9.02$ Å, $b \approx 5.21$ Å, and $c \approx 15.22$ Å, establishing a higher-order epitaxial relationship with $SrTiO_3$ (111) substrate with ~5.7% lattice mismatch (Supplementary Information Fig. S1). This enables the stabilization of two domains orthogonal to each other in-plane, as reflected by the observation of both $h0l$ and $0kl$ film peaks in the RSM measurement (Fig. 2d). The large degree of mismatch also explains the relaxed nature of the phase on $SrTiO_3$ (111) (Fig. 2d). The phase on $LaAlO_3$ ($p_2$) is hexagonal ($a \approx 5.52$ Å and $c \approx 22.15$ Å), with ~2.9% lattice mismatch with $LaAlO_3$ (111) substrate (in-plane substrate lattice constant of ~5.36 Å in the hexagonal setting under (111) orientation) (Fig. 2b). The unit cells of two phases on LSAT ($p_3$ and $p_4$, see Fig. 3b) are characterized to be hexagonal, epitaxially matched to the LSAT (111) substrate ($a \approx 5.47$ Å and $c \approx 8.90$ Å, with values derived from the convoluted XRD film peaks in Fig. 2a,c). Overall, these epitaxial relationships verify the role of epitaxy in the DAE synthesis process.

High-angle annular dark-field scanning transmission electron microscopy (HAADF-STEM) cross-sectional images of the three samples are shown in Fig. 3a-c. Compared to the sample on $LaAlO_3$ with relatively smooth substrate-film interface (Fig. 3a, top), the samples on LSAT and $SrTiO_3$ show a rougher substrate-film interface with isolated nucleation of the stabilized phases forming disconnected islands (Fig. 3b-c, top). This suggests a more vigorous diffusion process for the LSAT and $SrTiO_3$ samples, likely related to the tendency of Sr diffusion and surface segregation frequently observed in Sr-containing oxides.[30–32] We note that the four identified phases are the only crystalline phases observed via HAADF-STEM measurements, indicating that the other minority-phase peaks observed in Fig. 2a are potentially textured with directional coherence only in the sample out-of-plane direction. We also do not find any regions matching $p_1$ in cross-sectional HAADF-STEM images for the sample on LSAT (111), suggesting – together with the low-intensity peaks observed by XRD (Fig. 2a) – that any presence of that phase is likely textured and in the minority. Importantly, energy-dispersive X-ray spectroscopy (EDX) reveals that the phase stabilized on $SrTiO_3$ contains Sr, O, and Al, indicating that Ti diffused out while Al diffused in during the synthesis process (Fig. 3d-j). This confirms the simultaneous role of diffusion along with epitaxy in phase stabilization.

## Nature of the stabilized phases

The above structural characterizations reveal four distinct phases, with single-phase stabilization on $SrTiO_3$ ($p_1$) and $LaAlO_3$ ($p_2$) and the co-stabilization of two phases on LSAT ($p_3$ and $p_4$, with $p_3$ being the brighter of the two in HAADF-STEM) (Fig. 3a-c). Directly solving the crystal structures of the four stabilized phases is experimentally challenging given the small sample volume (e.g., effective film thickness of ~50 nm for the phase on $SrTiO_3$; see Supplementary Information Fig. S2). Therefore, we compare to inorganic structure databases (experimental[1–4] and

computational[4–10]) to identify if reported structures are compatible with the experimental observations of the four phases. Structures with which the number of distinct elements and element types are compatible with EDX (Fig. 3d-j and Supplementary Information Table S1) and cross-sectional STEM observations (Fig. 3a-c, Fig. 4, and Supplementary Information Fig. S3) are first collected. For each of these structures, crystallographic orientations that are compatible with the observed film peaks in XRD $\theta$-$2\theta$ symmetric scans and RSM measurements are then identified. These structures are finally verified against cross-sectional STEM images to determine the structure(s) that are compatible with all of the aforementioned structural characterizations (see Supplementary Information).

The results for the four phases are as follows. For $p_2$, through the structure identification process, $La_{0.9}Al_{11.95}O_{18.9}$ (COD-1001259)[1] is found to be compatible with the phase's structural characterization data (Fig. 4). In contrast, this process fails to identify any structure that is compatible with the structural characterization data of the three other phases. This suggests two possibilities. First, these three phases may correspond to reported structures that are not registered in the surveyed experimental[1–4] and computational[4–10] inorganic structure databases. Second, the three phases may be novel crystal structures that are neither experimentally realized before nor identified in computational materials searches that yielded the surveyed computational databases. Further discussions on the structural nature of these phases, comparison to literature, and computational investigations of these structures can be found in the Supplementary Information.

## Discussion

The present synthesis approach can be considered in comparison with the established metastable synthesis methods. DAE is similar to soft-chemistry methods in that interdiffusion of elements actively takes place and contributes to phase stabilization, but this process simultaneously provides metastability induced by epitaxy that is absent in conventional bulk soft-chemistry process. Further, DAE shares the characteristics of epitaxy in that the substrate provides an epitaxial platform to the phase being stabilized, but it additionally involves diffusion of elements that further enhances the kinetic nature of the synthesis environment. In this sense, the results in this work may be attributed to the unique synthesis condition established via simultaneous occurrence of diffusion and epitaxy. That is, the combination of metastable synthesis techniques may further enlarge the metastable synthesis window, thus promoting novel phase stabilization. This is potentially reflected in the infinite-layer nickelates, for example, where the two-step synthesis process combining epitaxy with soft-chemistry deintercalation[33,34] enables the stabilization of a highly metastable infinite-layer structure ~400 meV $atom^{-1}$ above the convex hull,[4] considerably larger than the typical metastability range of ~200 meV $atom^{-1}$.[35] Another instance of this is thermal laser epitaxy, where oxygen diffusion during high-temperature epitaxial growth of Ti has shown to stabilize metastable cubic TiO phase ~270 meV $atom^{-1}$ above the convex hull.[4,36]

From this perspective, the present work suggests a direction of doubly metastable synthesis for enlarging the experimental metastable stabilization window towards realization of new materials.

DAE operates in a straightforward apparatus of conventional bulk flux growth. This process is also highly tunable, adding substrate orientation and substrate elements as orthogonal parameters to the already rich synthesis phase space of flux growth. This motivates the further development of materials prediction methods that are sensitive to the metastable synthesis environment for enhanced predictive capabilities of phase stabilization under metastable conditions.[37–39]

## Methods

**Diffusion-Activated Epitaxy:** Single-side-polished substrates of $LaAlO_3$ (111), LSAT (111), and $SrTiO_3$ (111) were cut into ~2.5 × 5 × 0.5 $mm^3$ pieces and placed at the bottom of an alumina Canfield crucible, with the polished side facing up. Sn and Co powders in a desired stoichiometric ratio were then poured on top of the substrates into the crucible, with 4~5 g of total powder mass. The crucible was then vacuum-sealed ($P \leq 10^{-4}$ Pa) inside a quartz tube and introduced into a box furnace. The furnace was first heated to initial thermalizing temperature $T_i$ for 2 days at 100 °C/h, then cooled to synthesis temperature $T_s$ at 50 °C/day, staying at $T_s$ for 2 days. Finally, the quartz tube was then quickly extracted from the furnace and centrifuged to separate the liquid flux from the sample and quench-cool in air. For the $LaAlO_3$ and $SrTiO_3$ samples in Fig. 2 and Fig. 3, only one substrate piece was introduced inside the crucible, while for the LSAT sample in Fig. 2 and Fig. 3 all three substrate pieces (i.e., $LaAlO_3$, $SrTiO_3$, and LSAT) were placed into the crucible simultaneously. The majority-phase peaks on LSAT were also observed in a separate growth with just a single LSAT piece in the crucible (Supplementary Information Fig. S5). The detailed synthesis conditions of the three samples in Fig. 2 and Fig. 3 are shown in Supplementary Information Table S2.

**Structural Characterization:** XRD $\theta$-$2\theta$ symmetric scans and RSM measurements were performed with a monochromated Cu $K_{\alpha 1}$ source ($\lambda$ = 1.5406 Å). Cross-sectional STEM specimens were prepared by a standard focused ion beam (FIB) lift-out process on a Thermo Scientific Helios G4 UX FIB. HAADF-STEM images were acquired on an aberration-corrected Thermo Fisher Scientific Spectra 300 X-CFEG operated at 300 kV with a probe convergence semi-angle of 30 mrad and inner (outer) collection angles of 66 (200) mrad. iDPC-STEM images were collected on the same instrument using a segmented detector with inner and outer collection angles of 15 and 54 mrad, respectively. Segmented images were combined to obtain center-of-mass displacements along orthogonal directions and processed with the GetPotential function developed by ref. [40]. EDX mapping was performed on the same instrument equipped with a 1.8 steradian Dual-X detector. Contributions from Fe and Cu which arise from stray radiation or fluorescence of the lenses and sample grid have been removed from elemental atomic percent analysis. Additional measurements were also conducted on an FEI Titan Themis Cryo-S/TEM and a JEOL ARM300F.

## Acknowledgements

This work was supported by the Gordon and Betty Moore Foundation EPiQS Initiative, grant no. GBMF9070 (synthesis instrumentation and computation), the U.S. Department of Energy (DOE) Office of Science, Basic Energy Sciences, under award DE-SC0022028 (material development), the Office of Naval Research (ONR) under award N000142412407 (material analysis) and Army Research Office (ARO) grant no. W911NF-24-1-0234 (measurement technique development) to J.G.C. S.K. and H.J.K. acknowledge support from the MIT Energy Initiative Seed Fund (provided by Chuck Cahn). S.K., W.L., and H.J.K. acknowledge the MIT Office of Research Computing and Data, the MIT SuperCloud, and Lincoln Laboratory Supercomputing Center for providing high performance computing resources that have contributed to the research results reported in this work. S.K. acknowledges partial support from the ILJU Academy and Culture Foundation. Partial support was provided by the Molecular Sciences Software Institute (MolSSI) Software Fellowship Program (W.L.). The XRD characterizations were performed in the MIT.nano Characterization Facilities. This work made use of the Cornell Center for Materials Research shared instrumentation facility and Thermo Fisher Spectra 300 (Andromeda) STEM acquired with support by the NSF (DMR-2039380). Electron microscopy experiments were supported by the NSF Platform for Accelerated Realization, Analysis, and Discovery of Interface Materials (PARADIM) under Cooperative Agreement No. DMR-2039380. The FEI Titan Themis 300 was acquired through NSF-MRI-1429155, with additional support from Cornell University, the Weill Institute, and the Kavli Institute at Cornell.

## Author contributions

K.L. and J.G.C. conceived the project. K.L., Y.A., and M.F.D. fabricated the DAE samples. K.L. performed X-ray diffraction characterizations. B.H.G. and T.M. conducted electron microscopy measurements and EDX analyses. K.L. developed the structure identification process. S.K., W.L., and H.J.K. performed computations and structural refinements. K.L., S.K., T.S., H.J.K., and J.G.C. wrote the manuscript with input from all authors.

## Competing interests

The authors declare no competing interests.

## Data availability

The data that support the findings of this study are available from the corresponding authors upon request.

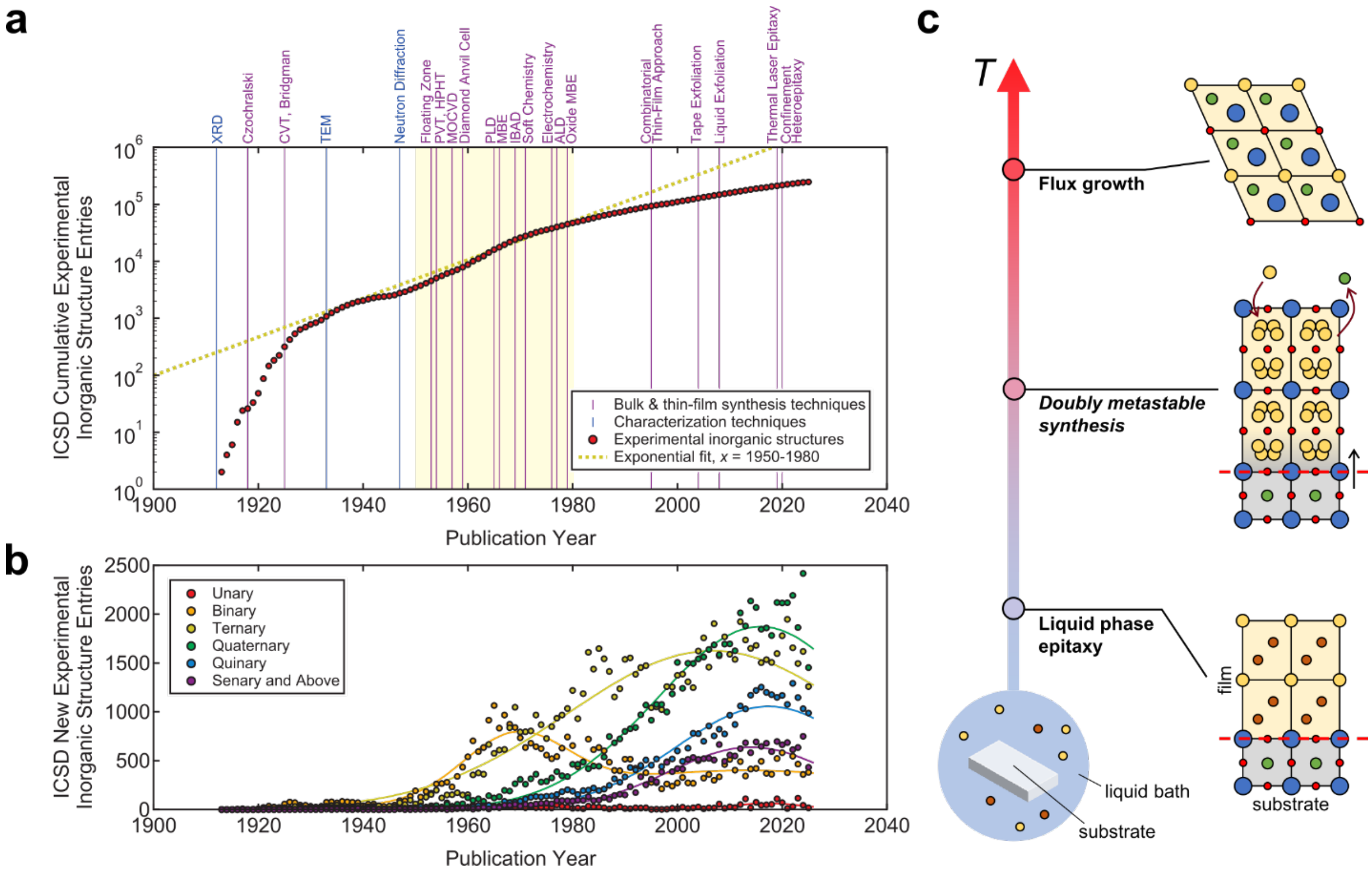

**Fig. 1 | Doubly metastable synthesis technique for accelerated materials discovery. a,** ICSD (release 2026.1)[2] cumulative experimental inorganic structure entries versus publication year (red markers), showing a doubling time of ~10 years (yellow dashed line) between ~1950 to ~1980 (yellow shaded region). Purple vertical lines indicate the invention years of a non-exhaustive list of synthesis techniques,[41–56,17,57–60,24,61,62] and the invention years of select characterization techniques are marked by blue vertical lines. **b,** ICSD new experimental inorganic structure entries versus publication year, distinguished by the number of distinct elements in the structure. Solid curves are Gaussian fits to the data. For unary and binary data points, two Gaussians were fitted for better qualitative agreement. **c,** Schematic of the doubly metastable synthesis technique. A higher-melting-point substrate is submerged in a liquid bath of elements then exposed to an intermediate temperature range at which the substrate surface bonds are activated to engage in diffusion reactions with the environment (brown arrows), while maintaining epitaxial energy (black arrow) produced by the inert bulk of the substrate. This provides a hybrid diffusion-epitaxy environment at the substrate surface.

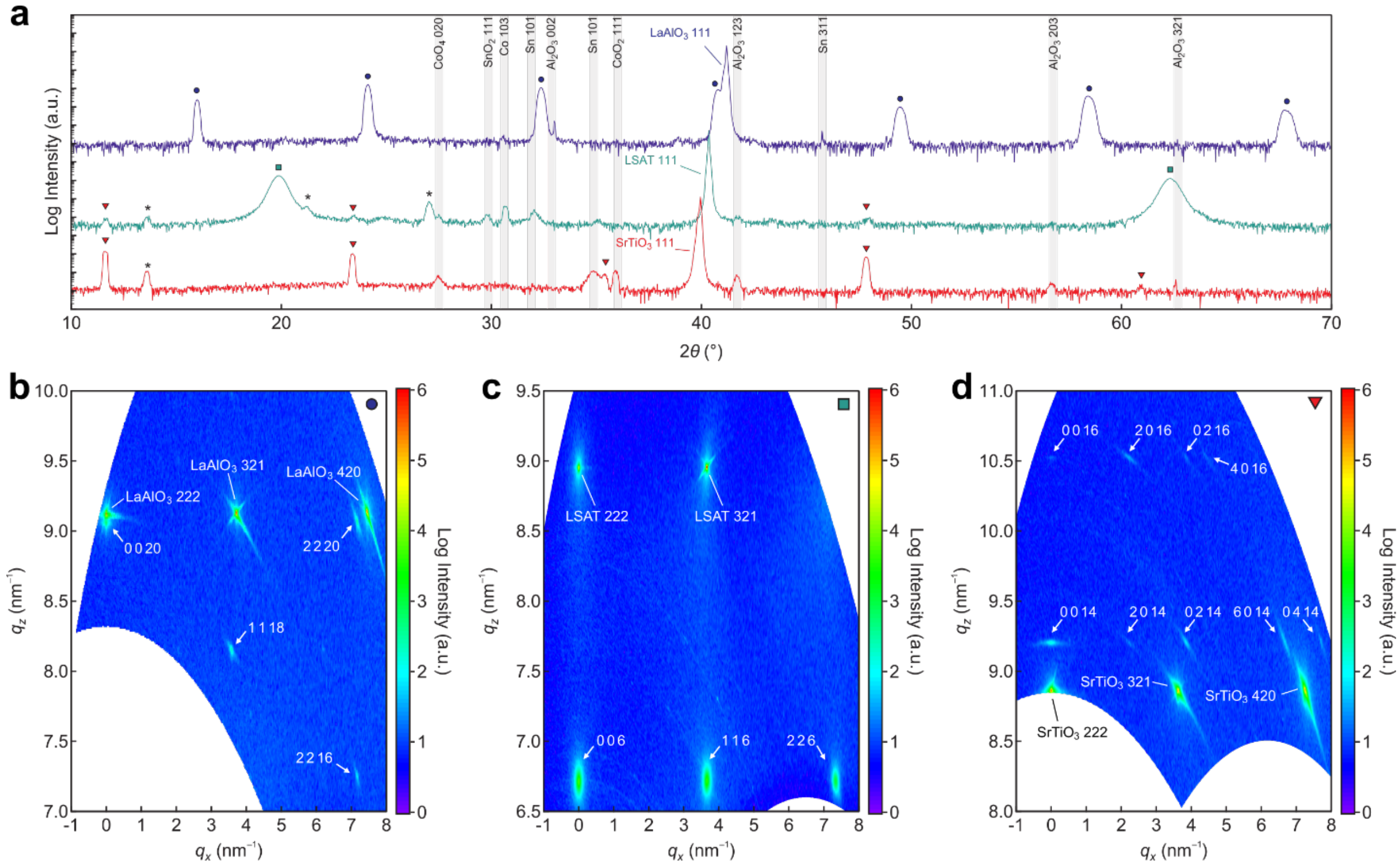

**Fig. 2 | X-Ray Diffraction Characterization of Phases Stabilized by DAE. a,** X-ray diffraction (XRD) $\theta$-$2\theta$ symmetric scan of the three samples stabilized on $LaAlO_3$ (111) (top blue), LSAT (111) (middle green), and $SrTiO_3$ (111) (bottom red) substrates. The curves are vertically offset for clarity. Three distinct sets of peaks corresponding to the stabilized phases are marked by red triangles, green squares, and blue circles, along with impurity phases from the flux in grey with plausible phase assignments labeled. Unidentifiable minority-phase peaks are indicated by black asterisks. The set of peaks marked by green squares are wide and high in intensity, suggestive of a convolution of two adjacent peaks. **b-d,** XRD reciprocal space map (RSM) measurements of the three samples in (**a**). The film peaks are indicated by white arrows, with their Miller indices labeled. The peak symbols in (**a**) corresponding to the majority phase for each sample are indicated at the upper right.

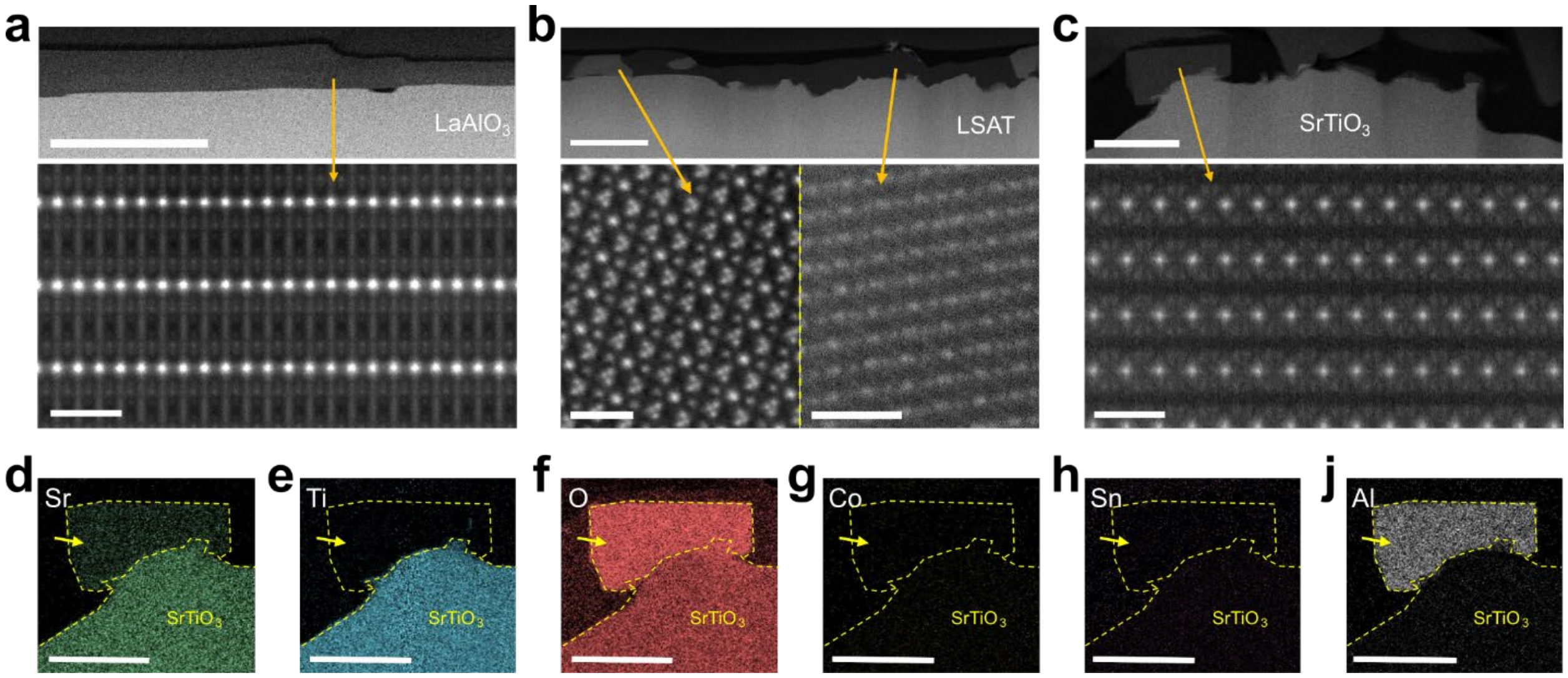

**Fig. 3 | Scanning Transmission Electron Microscopy Characterization of Phases Stabilized by DAE. a-c,** Wide-field-of-view (top, scale bar 1 μm) and magnified (bottom, scale bar 1 nm) high-angle annular dark-field scanning transmission electron microscopy (HAADF-STEM) cross-sectional images of the three samples. The LSAT sample shows a mixture of two distinct phases, both of which contribute to the green square peaks in Fig. 2a. **d-j,** Energy-dispersive X-ray spectroscopy (EDX) of the stabilized phase (yellow arrow) on $SrTiO_3$ (111), showing the presence of Sr, O, and Al and the absence of others. Scale bar 1 μm.

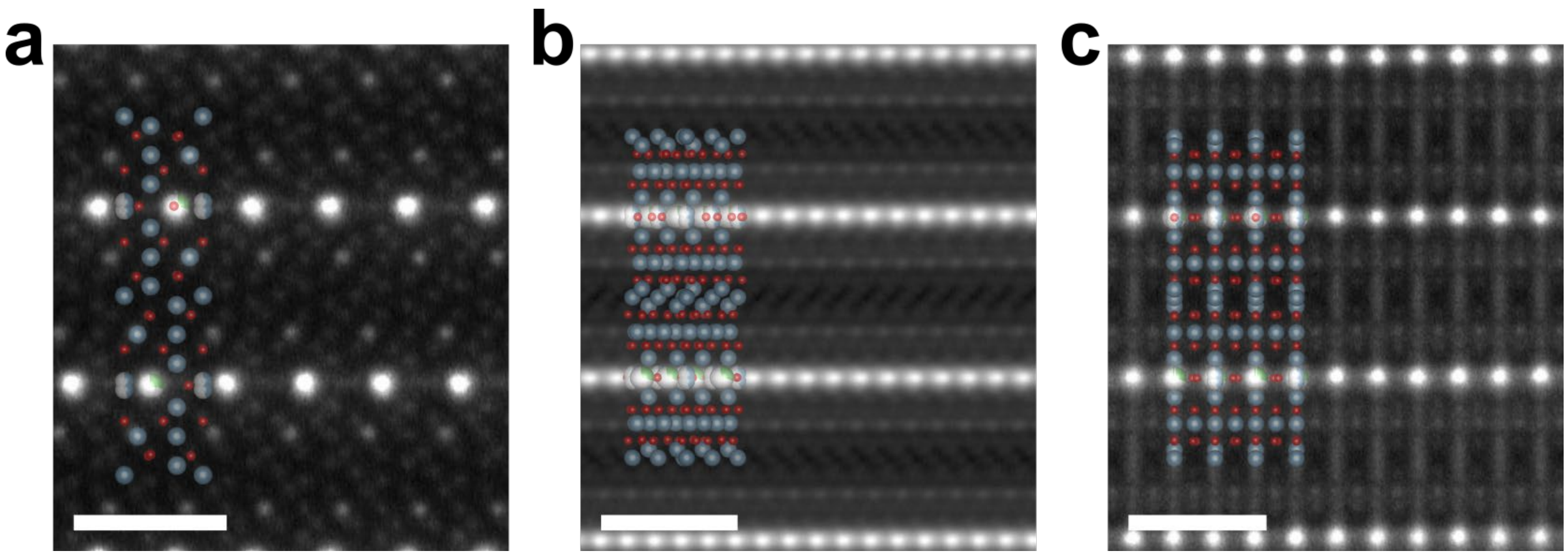

**Fig. 4 | Structure Identification Result for $p_2$.** Cross-sectional HAADF-STEM images of $p_2$ at three different zone axes, overlayed by $La_{0.9}Al_{11.95}O_{18.9}$ (001) at [100] (**a**), [28 9 0] (**b**), and [210] (**c**) zone axes. This structure is identified through the structure identification process to be consistent with $p_2$ and shows good agreement with the cross-sectional images of $p_2$, verifying the validity of this process. Scale bar 1 nm.